\documentstyle[12pt]{article}
\def\tr#1{{\rm tr}\kern-3pt\left[#1\right]}

\def\bea{\begin{eqnarray}}
\def\eea{\end{eqnarray}}
\def\nn{\nonumber}

\def\beq{\begin{equation}}
\def\eeq{\end{equation}}
\def\ba{\beq\new\begin{array}{c}}
\def\ea{\end{array}\eeq}
\def\be{\ba}
\def\ee{\ea}
\def\2{{1\over 2}}

\def\f{1\over}
\makeatletter
\newdimen\normalarrayskip              
\newdimen\minarrayskip                 
\normalarrayskip\baselineskip
\minarrayskip\jot
\newif\ifold             \oldtrue            \def\new{\oldfalse}
\def\arraymode{\ifold\relax\else\displaystyle\fi} 
\def\eqnumphantom{\phantom{(\theequation)}}     
\def\@arrayskip{\ifold\baselineskip\z@\lineskip\z@
     \else
     \baselineskip\minarrayskip\lineskip2\minarrayskip\fi}
\def\@arrayclassz{\ifcase \@lastchclass \@acolampacol \or
\@ampacol \or \or \or \@addamp \or
   \@acolampacol \or \@firstampfalse \@acol \fi
\edef\@preamble{\@preamble
  \ifcase \@chnum
     \hfil$\relax\arraymode\@sharp$\hfil
     \or $\relax\arraymode\@sharp$\hfil
     \or \hfil$\relax\arraymode\@sharp$\fi}}
\def\@array[#1]#2{\setbox\@arstrutbox=\hbox{\vrule
     height\arraystretch \ht\strutbox
     depth\arraystretch \dp\strutbox
     width\z@}\@mkpream{#2}\edef\@preamble{\halign \noexpand\@halignto
\bgroup \tabskip\z@ \@arstrut \@preamble \tabskip\z@ \cr}%
\let\@startpbox\@@startpbox \let\@endpbox\@@endpbox
  \if #1t\vtop \else \if#1b\vbox \else \vcenter \fi\fi
  \bgroup \let\par\relax
  \let\@sharp##\let\protect\relax
  \@arrayskip\@preamble}
\def\eqnarray{\stepcounter{equation}%
              \let\@currentlabel=\theequation
              \global\@eqnswtrue
              \global\@eqcnt\z@
              \tabskip\@centering
              \let\\=\@eqncr
              $$%
 \halign to \displaywidth\bgroup
    \eqnumphantom\@eqnsel\hskip\@centering
    $\displaystyle \tabskip\z@ {##}$%
    &\global\@eqcnt\@ne \hskip 2\arraycolsep
         $\displaystyle\arraymode{##}$\hfil
    &\global\@eqcnt\tw@ \hskip 2\arraycolsep
         $\displaystyle\tabskip\z@{##}$\hfil
         \tabskip\@centering
    &{##}\tabskip\z@\cr}
\makeatother

\def\Bf#1{\mbox{\boldmath $#1$}}

\def\balpha{{\Bf\alpha}}

\def\f{1\over }

\def\Bf#1{\mbox{\boldmath $#1$}}

\def\balpha{{\Bf\alpha}}

\begin{document}
\begin{titlepage}
\setcounter{footnote}0
\begin{center}
\hfill FIAN/TD-12/94\\
\hfill hep-th/9409190\\
\vspace{0.5in}
{\LARGE\bf Quantum Deformations of $\tau$-functions,}
\vspace{0.2in}
{\LARGE\bf
Bilinear Identities and Representation Theory}\footnote{Talk presented at the
Workshop on Symmetries and Integrability of Difference Equations, Montreal,
May, 1994}
\vspace{0.5in}
\\
{\Large A. Mironov\footnote{E-mail address:
mironov@lpi.ac.ru, mironov@nbivax.nbi.dk}}\\
\bigskip {\it Theory Department,  P. N. Lebedev Physics
Institute, Leninsky prospect, 53,\\ Moscow,~117924, Russia}\\{\large
and}\\{\it ITEP, Moscow, 117 259, Russia}
\end{center}

\bigskip
\bigskip
\bigskip

\centerline{\bf ABSTRACT}
\begin{quotation}
This paper is a brief review of recent results on the concept of
``generalized $\tau$-function'', defined as a generating function of
all the matrix elements in a given highest-weight
representation of a universal enveloping algebra ${\cal G}$.
Despite the differences
from the particular case of conventional $\tau$-functions of integrable
(KP and Toda lattice) hierarchies, these
generic $\tau$-functions also satisfy bilinear Hirota-like equations,
which can be deduced from manipulations with intertwining
operators.
The main example considered in details is
the case of quantum groups, when such $\tau$-``functions''
are not $c$-numbers but take their values in
non-commutative algebras (of functions on the quantum group $G$).
The paper contains only illustrative calculations for the simplest case of
the algebra $SL(2)$ and its quantum
counterpart $SL_q(2)$, as well as for the system of fundamental
representations of $SL(n)$.
\end{quotation}
\end{titlepage}
\clearpage

\newpage
\section{Introduction}
The key object in the theory of classical integrable equations is the
notion of $\tau$-function. This function allows one to transform non-linear
equations to bilinear homogeneous equations which are often called Hirota
equations. Indeed, there is even more important property of the
$\tau$-function and Hirota equations -- they can be easily extended to
an infinite set of equations satisfied by the same ($\tau$)-function of
(infinitely many) variables (times). The most effective way to deal with
this infinite set (and even to write down it)
is to encapsulate it into few generating identities which we will name
bilinear identities (BI). BI and $\tau$-functions are the main content
of the general approach to classical integrable systems as it has been
developed in the papers of Kyoto school \cite{DJKM}.

This paper, which is a review of the results obtained in collaboration with
A.Gerasimov, S.Kharchev, S.Khoroshkin, D.Lebedev, A.Morozov and L.Vinet
(see also \cite{R1,R2,R3,R4}),
demonstrates how this approach can be generalized and reformulated in
group theory terms so that the notion of $\tau$-function which
satisfies BI can be associated with an arbitrary group (and even quantum
group) and with any highest-weight representation. Indeed, this generalized
$\tau$-function is just defined as a generating function of all matrix
elements in a fixed representation. The standard KP (or Toda lattice)
hierarchy $\tau$-function \cite{DJKM} is associated with
the group $GL(\infty)$ and its fundamental representations. Different
KdV-type reductions are associated with the corresponding Kac-Moody algebras
with unit central charge.

In the present paper, in order to illustrate some new specific features of
the generalized $\tau$-functions we consider the case of quantized algebras,
mostly the simplest case of $SL_q(2)$ algebra. Another interesting example
of extension of the standard theory -- to Kac-Moody algebras with central
charge greater than 1, or even to multi-loop algebras - is out of scope of
the paper.

\section{$\tau$-function and bilinear identities}
\subsection{$\tau$-function}
Let us consider a universal enveloping algebra $U({\cal G})$ and introduce
a ``$\tau$-function'' for any Verma module $V$ of this algebra as a
generating function for all matrix elements
$\langle {\bf k} | g | {\bf n} \rangle_V$:
\be\label{tau}
\tau_V(t,\bar t|g) \equiv
\sum_{{{k_{\bf\alpha}\geq 0}\atop{n_{\bf\alpha}\geq 0}}}
\prod_{{\bf\alpha} > 0}
\frac{t_{\bf\alpha}^{k_{\bf\alpha}}}{[k_{\bf\alpha}]!}
\frac{\bar t_{\bf\alpha}^{n_{\bf\alpha}}}{[n_{\bf\alpha}]!}
\left._V \langle {\bf k}_{\bf\alpha} | g | {\bf n}_{\bf\alpha}
\rangle\right._V=\\
=\left._V
\langle {\bf 0} | \prod_{{\bf \alpha} > {\bf 0}} e_q(t_{{\bf\alpha}}
T_{{\bf\alpha}}) \ g \ \prod_{{\bf \alpha} > {\bf 0}}
e_q(\bar t_{{\bf \alpha}}T_{-{\bf \alpha}}) | {\bf 0} \rangle_V
\right. .
\ee
Here $[n] = \frac{q^n - q^{-n}}{q - q^{-1}}$,
$[n]! = [1][2]\ldots [n]$, $e_q(x) = \sum_{n\geq 0} \frac{x^n}{[n]!}$.
In the case of Lie algebras $q$-exponentials are substituted by
the ordinary ones. $T_{\pm{\bf\alpha}}$ are generators of positive/negative
maximal nilpotent subalgebras $N({\cal G})$ and
$\bar N({\cal G})$ of ${\cal G}$ with
suitably chosen ordering of positive roots ${\balpha}$, and
$t_{\bf\alpha}$, $\bar t_{\bf\alpha} = t_{-{\bf\alpha}}$ are the
associated ``time-variables''. Vacuum state is annihilated by
all the positive generators:
$T_{\bf\alpha} |{\bf 0}\rangle_V = 0$ for all ${\balpha} > 0$.
Verma module $V = \left\{
|{\bf n}_{\bf\alpha} \rangle_V=
\prod_{{\bf\alpha} > 0}
T_{-{\bf\alpha}}^{n_{\bf\alpha}}| {\bf 0}\rangle_V\right\}$
is formed by the action of all the generators $T_{-{\bf\alpha}}$
for all negative roots $-{\balpha}$ from maximal nilpotent
subalgebra $\bar N({\cal G})$.

Except for special circumstances
all the ${\balpha} \in N({\cal G})$ are involved,
and since not all the $T_{-{\bf\alpha}}$'s are commuting,
the so defined $\tau$-function has nothing to do
with Hamiltonian integrability (see \cite{Mor} for
detailed description of the specifics of $k=1$ Kac-Moody algebras in this
context). However, this appears to be the only property of conventional
$\tau$-functions which is not preserved by our general definition.
The "non-Cartanian" $\tau$-function and "non-Cartanian" hierarchy
still reduce to the standard integrable hierarchy when restricting onto
the fundamental representations which can be completely generated by
the commutative subalgebra of $N({\cal G})$.

\noindent
\underline{Example 1. $GL(\infty)$, $V=$fundamental representation}

Let us consider $GL(\infty)$
with Dynkin diagram infinite in both directions. Each vertex on the
diagram corresponds to a fundamental representation $F^{(n)}$ with
arbitrary fixed origin $n=0$ ($n$ can be both positive and negative).
This example admits free fermionic formulation \cite{DJKM} (see also
\cite{Mar,Mor,Mir}):
\be
\tau_n(t,\bar t|g) =
\langle F^{(n)} | e^{H\{t\}} g e^{\bar H\{\bar t\}}
|F^{(n)}\rangle
\label{fftau}
\ee
with
\be
H\{t\} = \sum_{n>0} t_nJ_{+n}, \ \ \ \bar H\{\bar t\} =
\sum_{n>0} \bar t_n J_{-n}, \ \ \ J_n =
\sum_{k=-\infty}^{\infty} \psi_k \psi^{\ast}_{k+n}, \ \ \
g = \exp \left(\sum_{m,n}{\cal
A}_{mn}\psi^{\ast}_n\psi_n\right),  \\
\{\psi_i,\psi_j^{\ast}\}=\delta_{ij},\ \ \{\psi_i,\psi_j\}=
\{\psi_i^{\ast},\psi_j^{\ast}\}=0;\ \
\psi^{\ast}_{k}|F^{(n)}\rangle = 0,
\ \ \hbox{for}\ \ k\ge n, \ \psi_{k}|F^{(n)}\rangle = 0 \ \ {\rm for}
\ \ k<n,
\label{ffdefs}
\ee
where two sets of currents $J_n$ with positive and negative $n$ give a
manifest realization of the commutative nilpotent subalgebras $N({\cal G})$
and $\bar N({\cal G})$ accordingly, and arbitrary generators of algebra
can be realized as bilinears $\psi_k\psi^{\ast}_{k+n}$.
In this example fermions intertwine different fundamental representations.

\noindent
\underline{Example 2. $SL(2)$}

In this case, the calculation is very simple for arbitrary highest weight
representation with spin $\lambda$:
\be\label{tau-sl2-clas}
\tau_\lambda=\left._\lambda\langle 0|e^{tT_-}ge^{\bar tT_+}
|0\rangle\right._\lambda = (a+b\bar t+ct+dt\bar t)^{2\lambda},
\ee
where the group element $g$ is parameterized by three parameters:
\be\label{g-clas}
g=e^{x_+T_+}e^{x_0T_0}e^{x_-T_-},
\ee
and
\be\label{abcd-clas}
a\equiv e^{{\f 2}x_0}+x_+x_-e^{-{\f 2}x_0},\ \ b\equiv x_+e^{-{\f 2}x_0},
\ \ c\equiv e^{-{\f 2}x_0}x_-,\ \ d\equiv e^{-{\f 2}x_0},
\ee
i.e.
\be\label{det-clas}
ad-bc=1.
\ee

\subsection{Vertex operators and bilinear identities}
\underline{Vertex operators}

1. The starting point is embedding of Verma module $\hat V$ into
the tensor product $V\otimes W$, where $W$ is some irreducible
finite-dimensional representation of ${\cal G}$.\footnote{In the
case of Affine algebra, one should use evaluation representation --
zero charge representation induced from finite-dimensional one -- see the
definition of vertex operator in \cite{FR,JM}.}
Once $V$ and $W$ are specified, there is only
finite number of choices for $\hat V$.

Now we define right vertex operator of the $W$-type
as homomorphism of ${\cal G}$-modules:
\be\label{inttw}
E_R: \ \ \hat V \longrightarrow V\otimes W.
\ee

This intertwining operator
can be explicitly continued to the whole representation once this is
constructed for its vacuum
(highest-weight) state:
\be\label{vacuum}
\hat V = \left\{ | {\bf n_{\bf\alpha}} \rangle_{\hat V} =
 \prod_{{\bf\alpha}>0}
 \left(\Delta (T_{-{\bf\alpha}}\right)^{n_{\bf\alpha}}
 | {\bf 0} \rangle_{\hat V} \right\},
\ee
where comultiplication $\Delta$ provides the action of ${\cal G}$ on
the tensor product of representations, and
\be\label{vacuum2}
|{\bf 0} \rangle_{\hat V} =
\left( \sum_{\{p_{\bf\alpha},i_{\bf\alpha}\}}
 A\{p_{\bf\alpha},i_{\bf\alpha}\}
 \left(\prod_{{\bf\alpha}>0}
 \left. (T_{-{\bf\alpha}}\right)^{p_{\bf\alpha}}\otimes
 \left. (T_{-{\bf\alpha}}\right)^{i_{\bf\alpha}}\right) \right)
| {\bf 0} \rangle_V \otimes | {\bf 0} \rangle_W.
\ee
For finite-dimensional $W$'s, this provides every
$| {\bf n_{\bf\alpha}} \rangle_{\hat V}$ in a form of {\it finite}
sums of states $| {\bf m_{\bf\alpha}} \rangle_{V}$ with
coefficients, taking values in elements of $W$.

2. The next step is to take another triple, defining left vertex
operator,
\be\label{inttw2}
E'_L: \ \ \hat V' \longrightarrow W' \otimes V',
\ee
such that the product $W\otimes W'$ contains {\it unit}
representation of ${\cal G}$.

\noindent
\underline{Bilinear identities in terms of universal enveloping algebra}

The derivation of BI consists of two steps. The first
one is to consider the projection to this unit representation
\be\label{proj}
\pi: \ \ W\otimes W' \longrightarrow I
\ee
explicitly provided by multiplication of any element
of $W\otimes W'$ by
\be\label{explproj}
\pi = \left._W \langle {\bf 0} | \otimes \left._{W'} \langle {\bf 0} |
\left( \sum_{\{i_{\bf\alpha},i'_{\bf\alpha}\}}
\pi\{i_{\bf\alpha},i'_{\bf\alpha}\}
 \left(\prod_{{\bf\alpha}>0}
 \left. (T_{+{\bf\alpha}}\right)^{i_{\bf\alpha}}\otimes
 \left. (T_{+{\bf\alpha}}\right)^{i'_{\bf\alpha}}\right) \right)
\right.\right.
\ee
Using this projection, one can build a new intertwining
operator
\be\label{Gamma}
\Gamma: \ \
\hat V \otimes \hat V' \stackrel{E_R\otimes E_L'}{\longrightarrow}
V \otimes W \otimes W' \otimes V'
\stackrel{I\otimes \pi \otimes I}{\longrightarrow} V \otimes V',
\ee
which possesses the property
\be\label{CRGamma}
\Gamma (g\otimes g) = (g\otimes g) \Gamma
\label{Ggg=ggG}
\ee
for any group element $g$ such that
\be\label{grel}
\Delta(g)=g\otimes g.
\ee

Put it differently, the space $W\otimes W'$ contains a canonical element of
pairing $w_i\otimes w^i$ which commutes with the action of $\Delta (g)$.
This means that the operator $\sum_i E_i\otimes E^i : V\otimes V'
\longrightarrow \hat V\otimes \hat V'$ ($E_i\equiv E(w_i),\ E^i\equiv
E(w^i)$) commutes with $\Delta (g)$.

Idenitity (\ref{Ggg=ggG}) is nothing but an algebraic form of BI.
To transform this to the differential (difference) form (the second step of
the derivation), one needs to use the second
line of definition (\ref{tau}) and to average identity (\ref{Ggg=ggG})
with the evolution exponentials over the universal enveloping algebra.
Then one gets
a BI for the averages like (\ref{tau}) (there are many equivalent
identities, in accordance with many possible choices of the states which
one averages over), but with additional
insertions of $E_i$'s and $E^i$'s. Using the commutation relations
of these intertwiners with the generators of the algebra, one can push
$E_i$'s out to the proper vacuums. This procedure of pushing can be immitated
by the action of some differential (difference) operators, and, as a result,
one gets, instead of averaged eq.(\ref{Ggg=ggG}), differential (difference)
BI.\footnote{In fact, in order to represent the result of
pushing by a differential or difference operator one needs to choose
properly the generating coefficients in definition (\ref{tau}). The
choice accepted in the paper is not unique.}

This second step of the derivation is very group-dependent,
and we illustrate it below in some concrete examples.

\noindent
\underline{Example 1. $GL(\infty)$, $V=$fundamental representation}

We will show in section 5.1 that the intertwining operators between
fundamental representations in this example are fermions (see example 1
in the previous subsection). The operator $\Gamma=\sum_i \psi_i\psi^i=
\sum_i \psi_i\psi^{\ast}_i$. Then, BI (\ref{Ggg=ggG}) gets
the form
\be\label{BIglinfty}
\sum_i \langle F^{(n+1)} | e^{H\{t\}} \psi_i g e^{\bar H\{\bar t\}}
|F^{(n)}\rangle \cdot \langle F^{(m-1)} | e^{H\{t'\}}
\psi_i^{\ast} g e^{\bar H\{\bar t'\}}
|F^{(m)}\rangle =\\
=\sum_i \langle F^{(n+1)} | e^{H\{t\}} g\psi_i
e^{\bar H\{\bar t\}}|F^{(n)}\rangle \cdot \langle F^{(m-1)} |
e^{H\{t'\}} g\psi_i^{\ast} e^{\bar H\{\bar t'\}}
|F^{(m)}\rangle,
\ee
where one averages (\ref{Ggg=ggG}) over the states
$\langle F^{(n+1)}|e^{H(t)}\otimes \langle F^{(m-1)}|e^{H(t')}$ and
$e^{\bar H(\bar t)}|F^{(n)}\rangle\otimes e^{\bar H(\bar t')}|F^{(m)}\rangle$.
One can rewrite (\ref{BIglinfty}) through the free fermion fields
$\psi (z)\equiv \sum_i \psi_i z^i$ and $\psi(z)^{\ast}\equiv\sum_i
\psi_i^{\ast}z^{-i}$:
\be
\oint_{\infty} {dz\over z}
\langle F^{(n+1)} | e^{H\{t\}} \psi(z) g e^{\bar H\{\bar t\}}
|F^{(n)}\rangle \cdot \langle F^{(m-1)} | e^{H\{t'\}}
\psi^{\ast}(z) g e^{\bar H\{\bar t'\}}
|F^{(m)}\rangle = \\=
\oint_0{dz\over z} \langle F^{(n+1)} | e^{H\{t\}} g\psi(z)
e^{\bar H\{\bar t\}}|F^{(n)}\rangle \cdot \langle F^{(m-1)} |
e^{H\{t'\}} g\psi^{\ast}(z) e^{\bar H\{\bar t'\}}
|F^{(m)}\rangle.
\ee
Now, using the relations
\be
\langle F^{(n)}|e^{H(t)}\psi(z)=
z^{n-1}\langle F^{(n-1)}|\exp [H(t_k- {1\over kz^k})]
\equiv z^{n-1}\hat X(z,t)\ \langle F^{(n-1)}|e^{H(t)},
\\
\langle F^{(n)}|e^{H(t)}\psi^{\ast}(z)=z^{-n}\langle F^{(n+1)}|\exp
[H(t_k+ {1\over kz^k})]
\equiv z^{-n}\hat X^{\ast}(z,t)\langle F^{(n+1)}|e^{H(t)},
\ee
where
\be
\hat X(z,t) = e^{\xi(z,t)} \; e^{-\xi(z^{-1},\tilde \partial_{t})},\ \
\hat X^{\ast}(z,t) = e^{-\xi(z,t)} \; e^{\xi(z^{-1},\tilde \partial_{t})},\ \
\xi(z,t)\equiv\sum_iz^it_i,\ \ \tilde \partial_{t_k}\equiv {\f k}
\partial_{t_k}
\ee
(and similarly for the right vacuum state),
one finally gets the integral form of BI:
\be
\oint_{\infty}dz z^{n-m}\hat X(z,t)\tau_n(t,\bar t)
\hat X^{\ast}(z,t')\tau_m(t',\bar t')=
\oint_0{dz\over z^2}z^{n-m}\hat X(z^{-1},\bar t)\tau_{n+1}(t,\bar t)
\hat X^{\ast}(z^{-1},\bar t')\tau_{m-1}(t',\bar t'),
\ee
which can be easily transformed to an infinite set of differential equations
by expanding to the degrees of time differences $t_i-t_i'$ etc. \cite{DJKM}.

\noindent
\underline{Bilinear identities in terms of algebra of functions}

Instead of averaging of (\ref{Ggg=ggG}) over the universal enveloping algebra
when deriving BI, one can work in terms of matrix
elements, i.e. at the dual language of the algebra of functions.
Indeed, let us take a matrix element of (\ref{Ggg=ggG})
between four states,
\be\label{CRGamma2}
\left._{V'} \langle k' | \left._V \langle k |
(g\otimes g) \Gamma | n \rangle_{\hat V}
 |n' \rangle_{\hat V'} \right.\right. =
\left._{V'} \langle k' | \left._V \langle k |
\Gamma (g\otimes g) | n \rangle_{\hat V}
|n' \rangle_{\hat V'} \right.\right..
\ee

The action of operator $\Gamma$ can be represented as
\be\label{me2}
\Gamma | n \rangle_{\hat\lambda} | n' \rangle_{\hat\lambda'}
= \sum_{l,l'} | l \rangle_\lambda | l' \rangle_{\lambda '}
\Gamma(l,l'| n,n'),
\ee
and (\ref{CRGamma2}) turns into
\be\label{me3}
\sum_{m,m'} \Gamma (k,k'| m,m')
\frac{|| k ||^2_\lambda
 || k' ||^2_{\lambda '}}
 {|| m ||^2_{\hat\lambda}
 || m' ||^2_{\hat\lambda '}}
\langle m | g | n \rangle_{\hat\lambda}
\langle m' | g | n' \rangle_{\hat\lambda '}
= \sum_{l,l'}
\langle k | g | l \rangle_\lambda
 \langle k' | g | l' \rangle_{\lambda '}
\Gamma(l,l'| n,n').
\label{bilimat}
\ee

In order to rewrite this as a difference equation, we
use the first line in definition (\ref{tau}) of $\tau$-function.
Then, one can write down the generating formula for equation
(\ref{bilimat}), using the manifest form (\ref{projA})-(\ref{projB})
of matrix elements $\Gamma(l,l'| n,n')$.
We illustrate this approach in the simplest
example of $SL_q(2)$ in section 3.1.

\section{$SL_q(2)$ case}
\subsection{Bilinear identities}
As an example of the technique developed in the previous section, we discuss
here BI and their solutions in the case of the simplest
quantum group $SL_q(2)$.

The notations: algebra $U_q(SL(2))$ has generators $T_+$,
$T_-$ and $T_0$ with commutation relations

\be
q^{T_0} T_\pm q^{-T_0} = q^{\pm 1}T_\pm, \ \
\phantom. [T_+,T_-] = \frac{q^{2T_0}-q^{-2T_0}}{q - q^{-1}},
\ee
and comultiplication

\be\label{coprod}
\Delta(T_\pm) = q^{T_0} \otimes T_\pm + T_\pm \otimes q^{-T_0}, \ \
\Delta(q^{T_0}) = q^{T_0}\otimes q^{T_0}.
\ee
Verma module $V_\lambda$ with highest weight $\lambda$ (not obligatory
half-integer), consists of the elements
\be\label{notation1}
| n \rangle_\lambda \equiv T_-^n|0 \rangle_\lambda, \ \ n\geq0,
\ee
such that
\be\label{notation}
T_- | n \rangle_\lambda = | n+1 \rangle_\lambda, \ \
T_0 | n \rangle_\lambda =
 (\lambda - n) | n \rangle_\lambda , \ \
T_+ | n \rangle_\lambda \equiv
b_n(\lambda) | n-1 \rangle_\lambda, \nn \\
b_n(\lambda) = [n][2\lambda + 1 - n], \ \ \
 [x] \equiv \frac{q^x - q^{-x}}{q - q^{-1}}, \ \
|| n ||^2_\lambda \equiv \left._\lambda\langle n | n \rangle\right._\lambda
= \frac{[n]!\ \Gamma_q(2\lambda +1)}{\Gamma_q(2\lambda +1-n)}
\stackrel{\lambda \in {\bf Z}/2}{=}
\frac{[2\lambda]![n]!}{[2\lambda -n]!}.
\ee
Now one could use the manifest commutation relations of intertwiners with
generators of $U_q(SL(2))$ to obtain BI along the line
of example 1, but instead, for the illustration of another approach, we
manifestly calculate matrix elements of the operator $\Gamma$.

Let us take for $W$ an irreducible
spin-$\2$ representation of $U_q(SL(2))$. Then $\hat
V=V_{\lambda\pm {\f 2}}$, $V=V_\lambda$.

In order to obtain matrix elements of $\Gamma$, one should
project the tensor product of two different $W$'s
onto singlet state
$S = |+\rangle |-\rangle - q|-\rangle|+\rangle$:

\be\label{qdet}
(A| + \rangle + B| - \rangle)\otimes
(| + \rangle C + | - \rangle D) \longrightarrow
AD-qBC.
\ee

With our choice of $W$ we can now consider
two different cases:

($A$) both $\hat
V=V_{\lambda-{\f 2}}$ and $\hat
V'=V_{\lambda'-{\f 2}}$, or

($B$) $\hat
V=V_{\lambda-{\f 2}}$ and $\hat
V'=V_{\lambda'+{\f 2}}$.

The result of calculation gives the following matrix elements of the
projection\footnote{Hereafter we omit
the symbol of tensor product from the notations of the states $|+\rangle
\otimes|0\rangle_\lambda$ etc.}:

\underline{Case A}:
\be\label{projA}
| n \rangle_{\lambda - \frac{1}{2}}
| n' \rangle_{\lambda ' - \frac{1}{2}}
\longrightarrow q^{\frac{n'-n-1}{2}} \left(
[n'-2\lambda']q^{\lambda '}
| n+1 \rangle_\lambda | n' \rangle_{\lambda '} -
[n-2\lambda]q^{-\lambda}
| n \rangle_\lambda | n'+1 \rangle_{\lambda '}\right).
\ee

\underline{Case B}:
\be\label{projB}
| n \rangle_{\lambda + \frac{1}{2}}
| n' \rangle_{\lambda ' - \frac{1}{2}}
\longrightarrow q^{\frac{n'-n-1}{2}} \left(
[n'-2\lambda']q^{\lambda '}
| n \rangle_\lambda | n' \rangle_{\lambda '} -
[n]q^{+\lambda +1}
| n-1 \rangle_\lambda | n'+1 \rangle_{\lambda '}\right).
\ee

Then, one can write down the generating formula for equation
(\ref{bilimat}), using the manifest form (\ref{projA})-(\ref{projB})
of matrix elements $\Gamma(l,l'| n,n')$:

\underline{Case A}:
\be\label{qeqA}
\sqrt{M_{\bar t}^- M_{\bar t'}^+}
\left( q^{\lambda '} D_{\bar t}^{(0)}
                   \bar t'D_{\bar t'}^{(2\lambda ')} -
  q^{-\lambda } \bar t D_{\bar t}^{(2\lambda)} D_{\bar t'}^{(0)}\right)
\tau_\lambda (t,\bar t | g)\tau_{\lambda '}(t',\bar t'| g) = \nn \\
= [2\lambda][2\lambda'] \sqrt{M_{t}^- M_{t'}^+}
\left( q^{-(\lambda+\frac{1}{2})}t' - q^{(\lambda' + \frac{1}{2})}t\right)
\tau_{\lambda - \frac{1}{2}}(t,\bar t | g)\tau_{\lambda ' - \frac{1}{2}}
(t',\bar t'| g).
\ee
Here
$
D_t^{(\alpha)} \equiv
\frac{q^{-\alpha}M^+_t - q^\alpha M^-_t}{(q - q^{-1})t}
$
and $M^{\pm}$ are multiplicative shift operators,
$M^{\pm}_tf(t)=f(q^{\pm 1}t)$.

\underline{Case B}:
\be\label{qeqB}
\sqrt{M_{\bar t}^- M_{\bar t'}^+}
\left( q^{\lambda '} \bar t' D_{\bar t'}^{(2\lambda ')} -
  q^{(\lambda +1)} \bar t D_{\bar t'}^{(0)}\right)
\tau_\lambda (t,\bar t | g)\tau_{\lambda '}(t',\bar t'| g) = \nn \\
= \frac{[2\lambda ']}{[2\lambda +1]} \sqrt{M_{t}^- M_{t'}^+}
\left( q^{\lambda '} tD_{t}^{(2\lambda +1)} -
  q^{\lambda} t' D_{t}^{(0)}\right)
\tau_{\lambda + \frac{1}{2}}(t,\bar t | g)\tau_{\lambda ' -
\frac{1}{2}}(t',\bar t'|g).
\ee

The classical limits of these equations are

\underline{Case A}:
\be\label{eqA}
\left(2\lambda \frac{\partial}{\partial \bar t'} -
    2\lambda' \frac{\partial}{\partial \bar t} +
 (\bar t' - \bar t)\frac{\partial^2}{\partial \bar t\partial \bar t'}
\right)
 \tau_\lambda (t,\bar t | g)\tau_{\lambda '}(t',\bar t'| g)
= 4\lambda\lambda '  (t' - t)
\tau_{\lambda - \frac{1}{2}}(t,\bar t | g)\tau_{\lambda ' - \frac{1}{2}}
(t',\bar t'| g) .
\ee

\underline{Case B}:
\be\label{eqB}
\phantom{fhg}\left[(\bar t'-\bar t){\partial\over\partial\bar t'}-2\lambda'
\right]
\tau_\lambda(t,\bar t|g)\tau_{\lambda '}(t',\bar t'|g)=
{2\lambda '\over 2\lambda +1}\left[(t-t'){\partial\over\partial t}-
2\lambda-1\right]\tau_{\lambda+{\f 2}}(t,\bar t|g)\tau_{\lambda '-
{\f 2}}(t',\bar t'|g).
\ee

\subsection{Solutions}
\underline{Classical limit and Liouville equation}

We begin with considering the solutions to the
classical BI. We only look at eq.(\ref{eqA}), as any
solution to this equation satisfies all other BI, say,
eq.(\ref{eqB}) or that obtained within a different choice of $V$, $\hat V$
and $W$. The general solution is 3-parametric one and certainly coincides
with the result (\ref{tau-sl2-clas}) of the direct calculation of section 2.1.

This solution (\ref{tau-sl2-clas}) at $\lambda={1\over 2}$ (fundamental
representation) seems to has a little to do with the solution to the
Liouville equation which is also sometimes associated with $SL(2)$. However,
there is a connection: though the Liouville equation has much more ample set
of solutions, eq.(\ref{eqA}) is contained in this set. Let us see how
additional limitations arise within our approach.

First of all, like all Hirota type equations, (\ref{eqA}) can be rewritten
as a (system of) ordinary differential equations, when expanded in
powers of $\epsilon = \frac{1}{2}(t - t')$ and
$\bar\epsilon = \frac{1}{2}(\bar t - \bar t')$. For example,
for $\lambda = \lambda'$ we obtain from (\ref{eqA}):
\be\label{expeq}
{\rm coefficient\ in\ front\ of\ }\epsilon:  \ \ \
\partial \tau_\lambda \bar\partial \tau_\lambda -
\tau_\lambda\partial\bar\partial\tau_\lambda =
2\lambda \tau_{\lambda - \2}^2; \nn \\
{\rm coefficient\ in\ front\ of\ }\bar\epsilon: \ \ \
2\lambda\tau_\lambda \bar\partial^2\tau_\lambda =
(2\lambda -1)(\bar\partial\tau_\lambda)^2; \nn \\
\ldots
\ee
If $\lambda = \frac{1}{2}$, the first one of these  is just
the Liouville equation:
\be\label{Liouv}
\partial \tau_{\2} \bar\partial \tau_{\2} -
\tau_{\2}\partial\bar\partial\tau_{\2} =\tau_0^2 = 1,
\ee
or
\be
\partial\bar\partial \phi = 2e^{\phi}, \ \
\tau_{\2} = e^{-\phi/2},
\ee
while the second one,
\be\label{Virasoro}
\bar\partial^2\tau_{\2} = 0,
\ee
is a very restrictive constraint.
Its role is to reduce the huge set of solutions to
the Liouville equation,
\be\label{Louvsol}
\tau_{{\f 2}}(t,\bar t|g)=(1+A(t)B(\bar t))\left[{\partial
A\over\partial t}{\partial B\over\partial\bar t}\right]^{-\2},
\ee
parametrized by two {\it arbitrary} functions $A(t)$ and $B(\bar t)$,
to the 3-parametric family (\ref{tau-sl2-clas}).
In the language of infinite-dimensional Grassmannian
there are infinitely many ways to embed $SL(2)$ group into
$GL(\infty)$ -- and all the cases correspond to solutions (\ref{Louvsol})
(with some $A(t)$ and $B(\bar t)$) to the $SL(2)$ reduced
Toda-lattice hierarchy (i.e. Liouville equation), - but constraint
(\ref{Virasoro})
specifies very concrete embedding: that in the left upper corner of
$GL(\infty)$ matrix and is associated with {\it linear} functions
$A(t)$ and $B(\bar t)$.

In terms of example 1 the general Liouville solution corresponds to
the matrix ${\cal A}_{mn}$ from (\ref{ffdefs}) of the form $\sum_{i=1,2}
F_n^{(i)}G_m^{(i)}$ with arbitrary coefficients $F^{(i)}_n$ and $G^{(i)}_m$
(i.e. to a matrix of the rank 2). Then, the corresponding element $g$ rotates
fermions in two-dimensional invariant subspace:
\be\label{at}
g\psi_i g^{-1}=\left(\int dt{e^{-tx}\over \sqrt{\partial A}}\int dxx^{i-1}
\right)\cdot\sum_k\left(\int d\bar t{e^{-\bar ty}\over\sqrt{\bar\partial B}}
\int dyy^{k-1}\right)\psi_k+\\+\left(\int dt{e^{-tx}A\over \sqrt{\partial A}}
\int dxx^{i-1}
\right)\cdot\sum_k\left(\int d\bar t{e^{-\bar ty}B\over\sqrt{\bar\partial B}}
\int dyy^{k-1}\right)\psi_k
\equiv f_i\Psi^{(1)}+g_i\Psi^{(2)}
\ee
in contrast to the general law \cite{DJKM}
\be
g\psi_i g^{-1}=\sum_j R_{ij}\psi_j.
\ee
Here the linear combinations of fermions $\Psi^{(1,2)}_i$ depend on
concrete choice of $g$, i.e. on the functions $A$ and $B$.
The whole variety of Liouville solutions is given
by different choices of the coefficients $F^{(i)}_n$, $G^{(i)}_m$ etc.
(they are connected with the moments of
Fourier components of the functions $A(t)$ and
$B(\bar t)$ like (\ref{at})), and different choices are related by
(outer) $GL(\infty)$ automorphisms of the $SL(2)$ system (see (\ref{at})).
Choosing only 2 first non-zero coefficients, one returns to the case of
the present paper.

\noindent
\underline{Quantum commutative $\tau$-function}

Now let us look at the solution to the quantum equation (\ref{qeqA}). One
can easily check that
\be\label{qsolA}
\tau_\lambda = [\alpha + {\f \alpha}t\bar t]^{2\lambda} \equiv
\sum_{i\geq 0} \frac{\Gamma_q(2\lambda +1)}
{\Gamma_q(2\lambda +1 - i)}\frac{\alpha^{2\lambda-2i}(t\bar t)^i}{[i]!}
\ee
does indeed satisfy (\ref{qeqA}), since
\be\label{1}
D_{t}^{(0)} [\alpha + {\f \alpha}t\bar t]^{2\lambda} = {\f \alpha}[2\lambda]
   [\alpha+ {\f \alpha}t\bar t]^{2\lambda -1}\bar t, \nn \\
tD_t^{(2\lambda)} [\alpha + {\f \alpha}t\bar t]^{2\lambda} = -\alpha
[2\lambda]
   [\alpha + {\f \alpha}t \bar t]^{2\lambda - 1}.
\ee
However, this is only 1-parametric solution, in contrast to the classical
case. This is due to the fact that, of all elements of $U_q(SL(2))$, the
only Cartan element has the proper comultiplication law (\ref{grel}), while
in the classical case there is the 3-parametric family of such elements
(\ref{g-clas}).

\noindent
\underline{Quantum non-commutative $\tau$-function}

The way to contruct whole family of solutions to the quantum BI
is to consider the non-commutative $\tau$-function. Indeed,
the first line of definition (\ref{tau}) implies that the
$\tau$-function takes its values in the algebra of functions on $SL_q(2)$,
i.e. is non-commutative quantity. For example, in fundamental representation
it is equal to
\be
\tau_{1\over 2}= \langle + | g | + \rangle  + \bar t \langle +
| g | - \rangle
+ t \langle - | g | + \rangle + t\bar t \langle - | g | - \rangle=
 a+b\bar t+ct+dt\bar t,
\ee
where generators $a,b,c,d$ of the algebra of functions $A(SL_q(2))$
are the entries of the matrix
\beq
{\cal T} = \left(
\begin{array}{cc}
 a b \\ c d
\end{array}
\right),\ \ \ \ \ ad-qbc=1
\eeq
with the commutation relations dictated by ${\cal T}{\cal T}{\cal R}=
{\cal R}{\cal T}{\cal T}$ equation \cite{FRT}
\be
ab = qba, \ \
ac = qca, \ \
bd = qdb, \ \
cd = qdc, \ \
bc = cb, \ \
ad - da = (q - q^{-1})bc.
\label{core}
\ee
In order to obtain this non-commutative $\tau$-function from the second line
of (\ref{tau}) and, simultaneously, to enlarge the number of group elements
satisfying condition (\ref{grel}), one needs to consider $g$ as an
element of the universal enveloping algebra given over {\it non-commutative}
ring instead of complex numbers. This ring is just $A_q(SL(2))$ (see the
next section).

In order to construct the non-commutative $\tau$-function for any
representation with spin $\lambda$, one can expand this represenation
to the representations with spins $\lambda-{1\over 2}$ and ${1\over 2}$:
\be
\phantom._{\lambda }\langle k | g | n \rangle_{\lambda} =
q^{-\frac{k+n}{2}}\left[
\phantom._{\lambda-{\f 2}} \langle k | g | n \rangle_{\lambda-{\f 2}}
\phantom. \langle + | g | + \rangle +
q^{\lambda }[n]\phantom._{\lambda-{\f 2}}\langle k | g | n-1
\rangle_{\lambda-{\f 2}} \langle + | g | - \rangle +\right.\\+ \left.
q^{\lambda }[k]\phantom._{\lambda-{\f 2}}\langle k-1 | g | n
\rangle_{\lambda-{\f 2}} \langle - | g | + \rangle +
q^{2\lambda}[k][n]\phantom._{\lambda-{\f 2}}\langle k-1 | g | n-1
\rangle_{\lambda-{\f 2}} \langle - | g | - \rangle \right].
\ee
Applying this procedure recursively, one gets:
\be
\tau_\lambda(t,\bar t|g) =
\tau_{\lambda - \2}(q^{-\2}t,q^{-\2}\bar t | g)
\tau_{\2}(q^{\lambda - \2}t,q^{\lambda - \2}\bar t | g) = \nn \\
\stackrel{{\rm if}\ \lambda \in \hbox{\bf Z}/2}{=}
\tau_{\2}(q^{\2 - \lambda}t, q^{\2 - \lambda}\bar t|g)
\tau_{\2}(q^{{3\over 2}-\lambda}t, q^{{3\over 2} - \lambda}\bar t | g) \ldots
\tau_{\2}(q^{\lambda - \2}t, q^{\lambda - \2}\bar t | g).
\ee

\section{Universal T-operator}
\subsection{Universal T-operator}
Let us manifestly describe the construction of "the group element" over
the non-commutative ring. More precisely, we describe such an element $g\in
U_q(G)\otimes A(G)$ of the
tensor product of the universal enveloping algebra $U_q(G)$ and its dual
$A(G)$ that
\be\label{Tcoprod}
\Delta_U(g)=g\otimes_{U} g \in A(G)\otimes U_q(G)\otimes U_q(G).
\ee
In order to construct this element \cite{FRT,FG,R4,R5}, we fix some basis
$T^{(\alpha)}$ in $U_q(G)$. Between $U_q(G)$ and $A(G)$
there is a non-generated pairing $<...>$. We
fix the basis $X^{(\beta)}$ in $A(G)$ orthonormal to $T^{(\alpha)}$
with respect to this pairing. Then, the sum
\be
\hbox{{\bf T}}\equiv\sum_{\alpha}X^{(\alpha)}\otimes T^{(\alpha)}\in A(G)
\otimes U_q(G)
\ee
is just the group element we need. It is called universal {\bf T}-matrix
(as it is intertwined by the universal ${\cal R}$-matrix).

In order to prove (\ref{Tcoprod}), one should remark that the matrices
$M^{\alpha\beta}_{\gamma}$ and $D^{\alpha}_{\beta\gamma}$
giving rise to multiplication and comultiplication laws in $U_q(G)$
respectively
\be
T^{(\alpha)}\cdot T^{(\beta)}\equiv M^{\alpha\beta}_{\gamma}T^{(\gamma)},\ \
\Delta(T^{(\alpha)})\equiv D^{\alpha}_{\beta\gamma}T^{(\beta)}\otimes
T^{(\gamma)}
\ee
induce, inversly, the comultiplication and multiplication in the dual algebra
$A(G)$:
\be\label{DM}
D^{\alpha}_{\beta\gamma}=\left<\Delta(T^{(\alpha)}),X^{(\beta)}\otimes
X^{(\gamma)}\right>\equiv \left<T^{(\alpha)},X^{(\beta)}\cdot
X^{(\gamma)}\right>,\\
M^{\alpha\beta}_{\gamma}=\left<T^{(\alpha)}T^{(\beta)},X^{(\gamma)}\right>=
\left<T^{(\alpha)}\otimes T^{(\beta)},\Delta(X^{(\gamma)})\right>.
\ee
Then,
\be
\Delta_U(\hbox{{\bf T}})=\sum_{\alpha}X^{(\alpha)}\otimes
\Delta_U(T^{(\alpha)})=
\sum_{\alpha,\beta,\gamma}D^{\alpha}_{\beta\gamma}X^{(\alpha)}\otimes
T^{(\beta)}\otimes T^{(\gamma)}=\sum_{\beta,\gamma}X^{(\beta)}X^{(\gamma)}
\otimes T^{(\beta)}\otimes T^{(\gamma)}=\hbox{{\bf T}}\otimes_U\hbox{{\bf T}}.
\ee
This is the first defining property of the universal {\bf T}-operator, which
coincides with the classical one. The second property, which allows one
to consider {\bf T} as generating the "true" group,
is the group composition law $g\cdot g'=g''$ given by the map:
\be
g\cdot g'\equiv \hbox{{\bf T}}\otimes_A
\hbox{{\bf T}}\in A(G)\otimes A(G)\otimes U_q(G)
\longrightarrow g''\in A(G)\otimes U_q(G).
\ee
This map is canonically given by the comultiplication and is
again the universal {\bf T}-operator:
\be
\hbox{{\bf T}}\otimes_A\hbox{{\bf T}}=\sum_{\alpha,\beta}X^{(\alpha)}\otimes
X^{(\beta)}\otimes T^{(\alpha)}T^{(\beta)}=\sum_{\alpha,\beta,\gamma}
M^{\gamma}_{\alpha,\beta}X^{(\alpha)}\otimes X^{(\beta)}\otimes T^{(\gamma)}=
\sum_{\alpha}\Delta(X^{(\alpha)})\otimes T^{(\alpha)},
\ee
i.e.
\be
g\equiv\hbox{{\bf T}}(X,T),\ \ g'\equiv \hbox{{\bf T}}(X',T),\ \
g''\equiv\hbox{{\bf T}}(X'',T),\\
\ \ X\equiv\{X^{(\alpha)}\otimes I\}\in A(G)\otimes I,
\ \ X'\equiv\{I\otimes X^{(\alpha)}\}\in I\otimes A(G),\ \
X''\equiv\{\Delta(X^{(\alpha)})\}\in A(G)\otimes A(G).
\ee

\subsection{Manifest construction of T-operator for $SL_q(2)$}

In order to get more compact formulas let us redefine the generators of
$U_q(SL(2))$ to obtain non-symmetric comultiplication law:
\be
T_+\longrightarrow T_+q^{-T_0},\ \ T_-\longrightarrow q^{T_0}T_-,\ \
\Delta(T_+)=I\otimes T_++T_+\otimes q^{-2T_0},\ \
\Delta(T_-)=T_-\otimes I+q^{2T_0}\otimes T_-.
\ee
{}From now on, we also change the definitions of $q$-numbers $[n]_q\equiv
{1-q^n\over 1-q}$ and, respectively, $q$-exponentials.

Now fix the basis $T^{(\alpha)}=T_+^iT_0^jT_-^k$ in $U_q(SL(2))$.
Then, from the coproduct of $T^{(\alpha)}$ one can calculate matrix
$D^{\alpha}_{\beta\gamma}$ (\ref{DM}) and manifestly calculate the
orthonormal basis of $X^{(\alpha)}$:
\be
X^{(\alpha)}={x_+^i\over [i]_{q^{-1}}!}{x_0^j\over j!}{x_-^k\over [k]_q!},
\ee
where the generating elements $x_{\pm},\ x_0$ satisfy the Borel Lie algebra
\be\label{xalgebra}
\phantom{.}[x_0,x_{\pm}]=(\ln q) x_{\pm},\ \ [x_+,x_-]=0.
\ee
Thus,
\be
\hbox{{\bf T}}=e_{q^{-1}}^{x_+T_+}e^{x_0T_0}e_q^{x_-T_-}.
\ee
This expression, indeed, very resembles element of the classical
Lie group and hints that the generators $T_{\alpha}$
in the definition of the $\tau$-function (\ref{tau}) should be rather taken
as elements of the extended algebra $U(G)\times A(G)$. Then, time variables
are nothing but parameters of $c$-number automorphisms of the dual algebra
$A(G)$. Say, in example (\ref{xalgebra}) $x_+$ and $x_-$ can be
multiplied by $c$-number factors, and the third $c$-number parameter can be
added to $x_0$.

\section{Fundamental representations of $SL(n)$}
\subsection{Intertwining operators}
This example is practically identical to example 1, however, we deal with it
in a more ``algebraic'' way to demonstrate some crucial points of the
approach.

There are as many as $r \equiv rank\ G = n-1$ fundamental
representations of $SL(n)$.
Let us begin with the simplest fundamental representation $F$ -
the $n$-plet, consisting of the states
\be\label{simplfrep}
\psi_i = T_-^{i-1}| 0 \rangle, \ \ i = 1,\ldots,n.
\ee
Here the distinguished generator $T_-$ is essentially a sum of
those for all the $r$ {\it simple} roots of $G$:
$T_- = \sum_{i=1}^r T_{-{\bf\alpha}_i}$.
Then all the other fundamental representations $F^{(k)}$ are
defined as skew powers of $F = F^{(1)}$:
\be\label{frep}
F^{(k)} = \left\{ \Psi^{(k)}_{i_1\ldots i_k} \sim
 \psi_{[i_1}\ldots \psi_{i_k]} \right\}
\ee
$F^{(k)}$ is essentially generated by the operators
\be\label{coprodFR}
R_k(T_-^i) \equiv T_-^i\otimes I \otimes \ldots \otimes I +
I\otimes T_-^i \otimes \ldots \otimes I +
I\otimes I \otimes \ldots \otimes T_-^i.
\ee
These operators commute with each other. It is clear that for
given $k$ exactly $k$ of them (with $i = 1,\ldots,k$) are
independent.

The intertwining operators which are of interest for us are
\be\label{inttwfrep}
I_{(k)}:\ \ F^{(k+1)} \longrightarrow F^{(k)} \otimes F, \ \ \
I^*_{(k)}:\ \ F^{(k-1)} \longrightarrow F^*\otimes F^{(k)},
\\ {\rm and} \ \ \
\Gamma_{k|k'}:\ \ F^{(k+1)}\otimes F^{(k'-1)} \longrightarrow
F^{(k)} \otimes F^{(k')}.
\ee
Here
\be\label{frep2}
F^* = F^{(r)} = \left\{\psi^i \sim \epsilon^{ii_1\ldots i_r}
  \psi_{[i_1}\ldots \psi_{i_r]} \right\}, \\
I_{(k)}:\ \ \Psi^{(k+1)}_{i_1\ldots i_{k+1}} =
            \Psi^{(k)}_{[i_1\ldots i_k}\psi^{\phantom{fgh}}_{i_{k+1}]}, \ \ \
I^*_{(k)}:\ \ \Psi^{(k-1)}_{i_1\ldots i_{k-1}} =
            \Psi^{(k)}_{i_1\ldots i_{k-1}i}\psi^i,
\ee
and $\Gamma_{k|k'}$ is constructed with the help of embedding
$I \longrightarrow F \otimes F^*$, induced by the pairing
$\psi_i \psi^i$: the basis in linear space $F^{(k+1)}\otimes
F^{(k'-1)}$, induced by $\Gamma_{k|k'}$ from that in
$F^{(k)}\otimes F^{(k')}$ is:
\be\label{prodfrep}
\Psi^{(k)}_{[i_1\ldots i_k}\Psi^{(k')}_{i_{k+1}]i'_1\ldots i'_{k'-1}}.
\ee
Operation $\Gamma$ can be now rewritten in terms of matrix elements
\be\label{gkdet}
g^{(k)}\left({{i_1\ldots i_k}\atop{j_1\ldots j_k}}\right) \equiv
\langle \Psi_{i_1\ldots i_k} | g | \Psi_{j_1\ldots j_k} \rangle=
\det_{1\leq a,b\leq k} g^{i_a}_{j_b}
\ee
as follows:
\be\label{gkgk}
g^{(k)}\left({{i_1\ldots i_k}\atop{[j_1\ldots j_k}}\right)
g^{(k')}\left({{i'_1\ldots i'_k}\atop
{j_{k+1}]j'_1\ldots j'_{k'-1}}}\right) =
g^{(k+1)}\left({{i_1\ldots i_k[i'_{k'}}
\atop{j_1\ldots j_{k+1}}}\right)
g^{(k'-1)}\left({{i'_1\ldots i'_{k'-1}]}\atop
{j'_1\ldots j'_{k'-1}}}\right)
\ee
This is the explicit expression for eq.(\ref{Ggg=ggG}) in the case of
fundamental representations, and it is certainly identically true for any
$g^{(k)}$ of form (\ref{gkdet}).

\subsection{$\tau$-function}
Now let us introduce time-variables and rewrite (\ref{gkgk}) in terms
of $\tau$-functions. We shall denote time variables through
$s_i, \bar s_i$, $i = 1,\ldots,r$ in order to emphasize their
difference from generic $t_{\bf\alpha}, \bar t_{\bf\alpha}$
labeled by all the positive roots ${\bf\alpha}$ of $G$. Note that in
order to have a closed system of equations we need to introduce all the
$r$ times $s_i$ for all $F^{(k)}$ (though $\tau^{(k)}$ actually depends
only on $k$ independent combinations of these).

Since the highest weight of representation $F^{(k)}$ is
identified as
\be
| {F^{(k)}}\rangle = |\Psi^{(k)}_{1\ldots k} \rangle,
\ee
we have:
\be\label{tau-k}
\tau_{k}(s,\bar s\ |\ g) =
\langle \Psi^{(k)}_{1\ldots k} |
\exp \left(\sum_i s_i R_k(T_+^i)\right)\ g \
\exp\left(\sum_i \bar s_iR_k(T_-^i)\right) |
\Psi^{(k)}_{1\ldots k} \rangle .
\ee

Now,
\be\label{70}
\exp\left(\sum_i s_i R_k(T^i)\right) =
\exp\left(R_k\left(\sum_i s_iT^i\right)\right)
= \left( \exp\left(\sum_i s_i T^i\right)\right)^{\otimes k} =
\left( \sum_j P_j(s)T^j\right)^{\otimes k},
\ee
where we used the definition of the Schur polynomials
\be\label{Schur}
\exp\left(\sum_i s_iz^i\right) = \sum_j P_j(s)z^j,
\ee
their essential property being:
\be\label{SS}
\partial_{s_i} P_j(s) = (\partial_{s_1})^i P_j(s) = P_{j-i}(s).
\ee
Because of (\ref{70}), we can rewrite the r.h.s. of (\ref{tau-k}) as
\be\label{detrep}
\tau_{k}(s,\bar s\ |\ g) =
\sum_{{i_1,\ldots,i_k}\atop{j_1,\ldots,j_k}}
P_{i_1}(s)\ldots P_{i_k}(s) \langle \Psi^{(k)}_{1+i_1,2+i_2,\ldots,k+i_k}
|\ g\ | \Psi^{(k)}_{1+j_1,2+j_2,\ldots,k+j_k} \rangle
P_{j_1}(\bar s)\ldots P_{j_k}(\bar s) =\\=
\det_{1\leq \alpha,\beta \leq k} H^\alpha_\beta(s,\bar s),
\ee
where
\be\label{H}
H^\alpha_\beta(s,\bar s) = \sum_{i,j} P_{i-\alpha}(s)g^i_j
P_{j-\beta}(\bar s).
\ee
This formula can be considered as including infinitely many
times $s_i$ and $\bar s_i$, and it is only due to the finiteness
of matrix $g^i_j \in SL(n)$ that $H$-matrix is additionally constrained
\be\label{excon}
\left(\frac{\partial}{\partial s_1}\right)^n H^\alpha_\beta = 0, \ \ \
\ldots \ \ \
\frac{\partial}{\partial s_i}H^\alpha_\beta = 0, \ \ {\rm for}\ \
i \geq n.
\ee
The characteristic property of $H^\alpha_\beta$ is that it satisfies
the following ``shift'' relations (see (\ref{SS})):
\be\label{der}
\frac{\partial}{\partial s_i}H^\alpha_\beta  = H^{\alpha +i}_\beta, \ \ \
\frac{\partial}{\partial \bar s_i} H^\alpha_\beta =
H^\alpha_{\beta + i}.
\ee
Expressions (\ref{detrep}), (\ref{H}) and (\ref{der})
are, of course, familiar from the theory of KP
and Toda hierarchies (see \cite{DJKM,Mor,Mar,Mir}
and references therein).

\subsection{Bilinear identities}
BI (\ref{gkgk}) can be easily
rewritten in terms of $H$-matrix: by convoluting them
with Schur polynomials. Let us denote
$H\left({\alpha_1\ldots\alpha_k}\atop{\beta_1\ldots \beta_k}\right)
= \det_{1\leq a,b \leq k} H^{\alpha_a}_{\beta_b}$. In accordance with
this notation
$\tau_{k} = H\left({1\ldots k}\atop{1\ldots k}\right)$, while the BI turns
into:
\be
H\left({\alpha_1\ldots \alpha_k}\atop{[\beta_1\ldots \beta_k}
\right)
H\left({\alpha'_k\alpha'_1\ldots \alpha'_{k-1}}\atop
{\beta_{k+1}]\beta'_1\ldots\beta_{k-1}'}\right) =
H\left({\alpha_1\ldots \alpha_k[\alpha'_k}\atop
{\beta_1\ldots\beta_k\beta_{k+1}}
\right)
H\left({\alpha'_1\ldots\alpha_{k-1}]'}\atop{\beta'_1\ldots\beta'_{k-1}}\right).
\ee
Just like original (\ref{gkgk}) these are merely matrix identities, valid for
any $H^\alpha_\beta$. However, after the switch from $g$ to $H$
we, first, essentially represented the equations in $n$-independent
form and, second, opened the possibility to rewrite them in terms
of time-derivatives.

For example, in the simplest case of
\be
\alpha_i = i, \ \ i = 1,\ldots, k'; \ \ \
\beta_i = i, \ \ i = 1,\ldots,k+1; \\
\alpha'_i = i, \ \ i = 1,\ldots,k-1,\ \ \alpha'_k = k+1; \ \ \
\beta'_i = i, \ \ i = 1,\ldots,k-1
\ee
we
get:
\be
H\left({1\ldots k}\atop{1\ldots k}\right)
H\left({k+1,1\ldots k-1}\atop{k+1,1\ldots k-1}\right) -
H\left({1\ldots k-1, k}\atop{1\ldots k-1, k+1}\right)
H\left({k+1,1\ldots k-1}\atop{k,1\ldots,k-1}\right) =\\=
H\left({1\ldots k+1}\atop{1\ldots k+1}\right)
H\left({1\ldots k-1}\atop{1\ldots k-1}\right)
\ee
(all other terms arising in the process of symmetrization vanish).
This in turn can be represented through $\tau$-functions:
\be\label{hirotafr}
\partial_1\bar\partial_1\tau_{k} \cdot \tau_{k} -
\bar\partial_1\tau_{k} \partial \tau_{k} =
\tau_{k+1}\tau_{k-1}.
\ee
This is the usual lowest Toda lattice equation. For finite $n$
the set of solutions is labeled by $g \in SL(n)$ as a result of
additional constraints (\ref{excon}).

\section{$\tau$-functions and Satsuma hierarchy}
\subsection{$\tau$-functions and representations of algebra of functions}
Non-commutative $\tau$-function in our definition is an element of algebra
of functions on the group. Therefore, one can fix different representations
of this algebra and look at ``the values'' of the corresponding
$\tau$-functions. The natural question arises how one can come to some
$c$-number functions in this way. The simplest $c$-number object is "double
generating function", which generates matrix elements of both representation
and co-representation of the universal enveloping algebra
(co-representation of the universal enveloping algebra = representation of
the algebra of functions). This double generating function depends on four
sets of time variables, and should satisfy BI with respect to both
representation and co-representation indices (therefore, this should be
some 4-dimensional system of equations).

Another $c$-number function is the $\tau$-function itself taken in trivial
co-representation. In order to understand better all this stuff let us
discuss what is the structure of co-representations in our usual example of
$SL_q(2)$.

Co-represenations are given by the corresponding representations of algebra
(\ref{xalgebra}). This is a Borel algebra, and, therefore, it has no
finite-dimensional non-trivial irreps \cite{Zhe}. All finite-dimensional
representations are reducible, but not completely reducible. Let us write
down manifestly the connection (bosonization) of the standard generators of
$A(SL_q(2))$ (\ref{core}) in terms of algebra (\ref{xalgebra}):
\be
a=e^{{\f 2}x_0}+x_+x_-e^{-{\f 2}x_0},\ \ b= x_+e^{-{\f 2}x_0},
\ \ c= e^{-{\f 2}x_0}x_-,\ \ d= e^{-{\f 2}x_0}.
\ee
This expression coincides with (\ref{abcd-clas}), the commutation
relations between $x$'s being only different.

Thus, there are only two irreps:
the trivial one, which is given by $x_+=x_-=0$, i.e. $ad=1,\ b=c=0$, and
the infinite dimensional irrep, given manifestly by the action on the
basis $\{e_k\}_{k\ge 0}$ \cite{Soi}
\be
ae_k=(1-q^{2k})^{\2}e_{k-1}\ (ae_0=0),\ \ de_k=(1-q^{2k+2})^{\2}e_{k+1},\ \
ce_k=\theta q^ke_k,\ \ be_k= -\theta^{-1}q^{k+1}e_k.
\ee

This picture can be easily generalized to other quantum groups (of the rank
$r$), as the corresponding algebras of $x$'s are always Borel algebras.
Therefore, there are only trivial and infinite-dimensional (both
$r$-parametric) irreps of the algebras of functions in these cases \cite{Soi}.

\noindent
\underline{$\tau$-function in trivial representation}

It was already mentioned that, taken in trivial representation of the algebra
of functions, $\tau$-function presents another example of $c$-number
function. Hence, this particular example is of great importance. Let us
first consider the simplest case of $SL_q(2)$. Then, there is no analog of
the determinant expression (\ref{detrep}) for the $\tau$-function. Indeed,
let us introduce (hereafter we denote $D\equiv D^{(0)}$):
\be
H^1_1 = \tau_F = a + b\bar t + ct + dt\bar t.
\ee
If
\be
H^1_2 = D_{\bar t}H^1_1 = b+dt, \ \ \
H^2_1 = D_t H^1_1 = c+d\bar t, \ \ \
H^2_2 = D_{\bar t}D_t H^1_1 = d,
\ee
we see that $H^a_b$ is actually not lying in $SL_q(2)$
(for example,
$H^1_2 H^2_1 \neq H^2_1 H^1_2$), i.e. the matrix consisting of
the $\tau_F$ and its derivatives, despite these are all
elements of $A(G)$, does not longer belong to $G_q$. Thus, it is not
reasonable to consider ${\rm det_q}H$ (or the definition of $H$ should
be somehow modified). Instead the appropriate
formula for the case of $SL_q(2)$ looks like
\be\label{sl2hirota}
\tau_{F^{(2)}} = {\rm det}_q g = 1=
H^1_1 H^2_2 - q H^1_2 M^-_{\bar t} H^2_1 =
\tau_F D_tD_{\bar t}\tau_F - q D_{\bar t}\tau_F M^-_tD_t\tau_F.
\ee
Now one can take trivial representation of $A(SL_q(2))$: $ad=1,\ b=c=0$
and obtain that the equation (this was first proposed by Satsuma et al.
\cite{Sat})
\be\label{hirotadif}
\tau_F D_tD_{\bar t}\tau_F - D_{\bar t}\tau_F D_t\tau_F=1
\ee
is correct. This equation looks like (\ref{hirotafr}) and has the form
independent of the concrete $SL_q(2)$ algebra. Moreover, equation
(\ref{sl2hirota}) seems not to be generalizable to $SL_q(n)$-case, in
contrast to (\ref{hirotadif}) (being applied only to $\tau$-functions in
trivial representations).

Another argument in favor of equation (\ref{hirotadif}) is that it
immediately leads to determinants of $q$-Schur polynomials. On the other
hand, the appearance of these is rather natural in the trivial
representations.
Indeed, let us consider unit element $g$ (which corresponds to taking
trivial representation of A(G)). Then, for $SL_q(n)$ algebra, one obtains
by direct calculation \cite{R1}
\be
\langle k,0,\ldots,0|e_{q^{-1}}^{s_2T_{12}}e_{q^{-1}}^{s_3T_{13}}\ldots
e_{q^{-1}}^{s_nT_{1n}}\times e_q^{\bar s_2T_{21}}e_q^{\bar s_3T_{31}}\ldots
e_q^{\bar s_nT_{n1}}|k,0,\ldots,0\rangle=P^{(q)}_k(s\bar s),
\ee
where $\langle k,0,\ldots,0|$ denotes the symmetric product of $k$
simplest funcdamental representations.

Because of all these reasons we are led to consider equation
(\ref{hirotadif}) in more details.

\subsection{Satsuma difference hierarchy}
\underline{Satsuma hierarchy from Toda lattice hierarchy}

Now we demonstrate that the {\it difference} (Satsuma) equation
(\ref{hirotadif}) can be obtained in the framework of the standard
{\it differential} Toda
lattice ($GL(\infty)$) hierarchy by the redefiniton of time flows \cite{R2}.
Indeed, Eq.(\ref{hirotadif}) is a corollary of {\it two} statements: the basic
identity (\ref{gkgk}) and the particular definition (\ref{tau}), which in
this case implies (\ref{H}) with $P$'s being ordinary
Schur polynomials (\ref{Schur}). At least, in this simple situation
(of fundamental representations of $SL(n)$) one could define
$\tau$-function not by eq.(\ref{tau}), but just by eq.(\ref{detrep}), with
\be
H^\alpha_\beta(s,\bar s) \longrightarrow
{\cal H}^\alpha_\beta(s,\bar s) = \sum_{i,j} {\cal P}_{i-\alpha}(s)\
g^i_j\ {\cal P}_{j-\beta}(\bar s)
\ee
with {\it any} set of independent functions (not even polynomials)
${\cal P}_\alpha$. Such
\be
\tau_{k}^{({\cal P})} = \det_{1\leq \alpha,\beta \leq k}
{\cal H}^\alpha_\beta
\ee
still remains a generating function for all matrix elements of $G=SL(n)$
in representation $F^{(k)}$. This freedom should be kept in
mind when dealing with ``generalized $\tau$-functions''. As a simple
example, one can take ${\cal P}_\alpha(s)$ to be $q$-Schur polynomials,
\be
\prod_i e_{q^i}(s_iz^i) = \sum_j P^{(q)}_j(s)z^j,
\ee
which satisfy
\be
D_{s_i} P^{(q)}_j(s) = (D_{s_1})^i P^{(q)}_j(s) = P^{(q)}_{j-i}(s).
\ee
Then instead of (\ref{der})
we would have:
\be
D_{s_i}{\cal H}^\alpha_\beta = {\cal H}^{\alpha +i}_\beta, \ \
D_{\bar s_i}{\cal H}^\alpha_\beta = {\cal H}^\alpha_{\beta+i}
\ee
and
\be
\tau^{(P^{(q)})}_{k} (s,\bar s | g) =
\det_{1\leq \alpha,\beta \leq k} D_{s_1}^{\alpha -1}
D_{\bar s_1}^{\beta -1} {\cal H}^1_1(s,\bar s).
\ee
So defined $\tau$-function satisfies (\ref{hirotadif}) \cite{Sat,R2}:
\be
\tau_{k}\cdot D_{s_1}D_{\bar s_1}\tau_{k}-
D_{s_1}\tau_{k}\cdot D_{\bar
s_1}\tau_{k}=\tau_{k-1}\cdot M^+_{s_1}M^+_{\bar s_1}\tau_{k+1},\\
\ldots\ \ \
\ee
where $\ldots$ means other equations of the (Satsuma) hierarchy.
The simplest way to prove this formula is to rewrite $\tau$-function using
\be
\det D_{s_1}^i D_{\bar s_1}^j H=
q^{-(N-1)(N-2)}(1-q)^{N(N-1)}(t\bar t)^{{N(N-1)\over 2}}
\det_{0\le i,j < N}  (M^+_{s_1})^i (M^+_{\bar s_1})^j H
\label{taudetdif}
\ee
and then to apply to this $\tau$-function the Jacobi identity. The Jacobi
identity is a particular ($p=2$) case of the general identity for the
minors of any matrix,
\be
\sum_{i_p}H_{ri_p}\hat H_{i_1\ldots i_p|j_1\ldots j_p} = \frac{1}{p!}\sum_P
(-)^P
\hat H_{i_1\ldots i_{p-1}|j_{P(1)}\ldots j_{P(p-1)}} \delta_{rj_{P(p)}},
\ee
where the sum on the r.h.s.\ is over all permutations of the $p$
indices and $\hat H_{i_1\ldots i_p|j_1\ldots j_p}$ denotes the
determinant (minor) of the matrix, which is obtained from
$H_{ij}$ by removing the rows $i_1\ldots i_p$ and the columns
$j_1\ldots j_p$. Using the fact that $(H^{-1})_{ij} = \hat
H_{i|j}/\hat H$, this identity can be rewritten as
\be
\hat H \hat H_{i_1\ldots i_p|j_1\ldots j_p} =
\left(\frac{1}{p!}\right)^2\sum_{P,P'} (-)^P(-)^{P'}
\hat H_{i_{P'(1)}\ldots i_{P'(p-1)}|j_{P(1)}\ldots j_{P(p-1)}}
\delta_{i_{P'(p)}|j_{P(p)}}.
\ee

\noindent
\underline{Fermionic approach to Satsuma hierarchy}

Let us look at the fermionic description of the Satsuma hierarchy. As far as
this hierarchy is obtained from the Toda lattice hierarchy (see example 1)
by the redefiniton of time variables, one can immediately use the
$\tau$-function of the Toda lattice hierarchy, and just substitute new times.
Indeed,
\be
\prod_{k=1}^\infty e_{q^k}(s_k z^k) = \prod_{k=1}^\infty e^{t_kz^k},
\ee
provided the $t$'s are expressed in terms of the $s$'s according to
\be
\sum_{k=1}^\infty t_kz^k = \sum_{n,k=1}^\infty
\frac{s_k^n(1-q_k)^n}{n(1-q_k^n)} z^{nk}.
\label{tverT}
\ee
Thus
\be
P^{(q)}_k(s) = P_k(t).
\ee

Because of this, $\tau$-function can be represented as
\be
\tau_{n}(s,\bar s|g) = \tau_{n}(t,\bar t|g)\ \stackrel{(\ref{fftau})}{=} \
\langle F^{(n)} | e^{H\{t\}} g e^{\bar H\{\bar t\}}
|F^{(n)}\rangle
\ee
and
\be\label{49}
H\{t\} = \sum_{n>0} t_nJ_{+n} \ \stackrel{(\ref{tverT})}{=} \
\sum_{n,k=0}^\infty \frac{s_k^n(1-q_k)^n}{n(1-q_k^n)} J_{+nk}, \nn \\
\bar H\{\bar t\} = \sum_{n>0} \bar t_n J_{-n} =
\sum_{n,k=0}^\infty \frac{\bar s_k^n(1-q_k)^n}{n(1-q_k^n)} J_{-nk}.
\ee

The Satsuma $\tau$-function can be also considered as some
Miwa transformed Toda $\tau$-function \cite{R2}. Indeed, the general Miwa
transformation of times would be $t_n={1\over n}\sum_i\lambda^{-i}$ with
sum running over generally infinite set of integer numbers. Then, using
formulas analogous to
\be
t_k = \frac{1}{k}\frac{((1-q)s_1)^k}{1-q^k} =
\frac{1}{k}\sum_{l \ge 0} \left( (1-q)q^l s_1\right)^k,
\label{semperMitr}
\ee
one gets the following set of Miwa variables leading to the Satsuma hierarchy
\be\label{Miwa}
\left\{ \left.e^{2\pi i a/k}\lambda_k q_k^{-l/k}   \right| a=0,\ldots k-1;
\ \ l\geq 0\right\}, \ \ \
\lambda_k = \left( (1-q_k)s_k \right)^{-1/k}.
\ee

Thus, after Miwa transformation of the Toda
lattice hierarchy, with the specific set of Miwa variables (\ref{Miwa})
one gets the Satsuma hierarchy.

\section{Concluding remarks}
In this paper we have introduced the notion of generalized $\tau$-function
and demonstrated that it satisfied a set of BI. We also have discussed
non-commutative $\tau$-functions arising in the framework of quantum groups.

The generalized $\tau$-function, with associated BI, is supposed to play a
very important role in applications. First, it gives a tool to investigate
WZW theories with level greater than 1 in integrable treatment.
Further development in the same direction might be dealing with 2-loop,
3-loop etc. algebras. Another application of the approach advocated in the
present paper is the case of quantum deformed algebras. This case has to have
much to do with careful quantization of the Liouville theory \cite{Weight},
and also might shed a light on the famous fact observed in quantum
integrable systems, whose correlators (more precisely, their generating
functional) satisfy some classical integrable equations
\cite{BIKSL}. This phenomenon is similar to that observed in the
connection of theories of $2d$ gravity and matrix models \cite{Mir}.

\section*{Acknowledgements}
I am indebted to my collaborators A.Gerasimov, S.Kharchev, S.Khoroshkin,
D.Lebedev, A.Morozov and L.Vinet. I also acknowledge A.Zabrodin for
numerous discussions. I would like to thank the organizers of the
Workshop on Symmetries and Integrability of Difference Equations for
the invitation and kind hospitality.
The work is partially supported by grant 93-02-03379 of the Russian
Foundation of Fundamental Research and by grant MGK000 of the International
Soros Foundation.

\end{document}